\documentstyle[12pt]{article}

\def\e{\epsilon}
\def\ra{\rightarrow}
\def\a{\alpha}
\def\b{\beta}
\def\n{\eta}
\def\ga{\gamma}
\def\d{\delta}

\def\cg{\cal G}

\def\pr{\partial }

B

\def\l{{\lambda }}
\newcommand{\be}{\begin{equation}} \newcommand{\ee}{\end{equation}}
\newcommand{\bea}{\begin{eqnarray}}\newcommand{\eea}{\end{eqnarray}}
\begin{document}
\baselineskip= 24 truept
\begin{titlepage}

\title { Planckian Scattering of Non-abelian \\
Gauge Particles }

\author{Supriya Kar $^{1}$ and Jnanadeva Maharana$^{2}$ \\
Institute of Physics, Bhubaneswar-751 005, India.}

\footnotetext[1]{e-mail:supriya@iopb.ernet.in}
\footnotetext[2]{e-mail:maharana@iopb.ernet.in}

\date{}
\maketitle

\thispagestyle{empty}

\vskip .6in
\begin{abstract}

\vskip .2in

We present a systematic study of high energy
scattering of non-abelian gauge particles in (3+1) dimensional
Einstein gravity using semi-classical techniques of Verlinde and
Verlinde. It is shown that the BRST gauge invariance of the
Yang-Mills action in presence of quantum gravity at Planckian
energy regime is maintained and the vertex operator is invariant
under the BRST transformations. The presence of gravitational
shock wave describing the gauge particles is discussed in the
resulting (3+1) dimensional effective theory of Yang-Mills
gravity.

\end{abstract}

\bigskip
\flushright{IP-BBSR/94-51}

\vfil
\end{titlepage}

\section{Introduction}

\hspace{.1in} In recent years the high energy scattering in quantum gravity
at Planckian energy has attracted  considerable attention. Several
attempts have been undertaken to understand the phenomena and it is
argued that a clear understanding of such processes will have to
account for the effects of quantum gravity. It is felt that a
non-perturbative treatment of the quantum gravity will be essential
to envisage collisions at the Planckian energies. On the other
hand, now there are convincing results which show that the
semi-classical techniques$^{1,2}$ can be used to compute the
Planckian scattering amplitudes when the momentum transfer
during the scattering  is small compared to the Planck scale. In
this kinematical regime a simple physical picture emerges, when
one considers the collision of two neutral particles$^1$, of
masses negligible compared to the Planck mass, $M_P$, at centre
of mass energy of the order of  $M_P$. One of the in-coming
particle finds a shock wave$^1$ approaching and the effect of
this shock wave is to add a phase to the wave function. The
scattering amplitude thus computed corresponds to the one
in eikonal approximation to leading order$^{3,4}$. This
scenario was first realised by 't Hooft$^1$. Furthermore it is
pointed out in Ref. 5 that the effective action desribing the
scattering processes has one piece which is in close resemblance
with the string action with an imaginary string tension and has
a vertex operator which is analogous to the one encounters in
string theory$^6$. In this context, it is worthwhile to note that the
scattering processes at Planckian energy have been envisaged in the
frame-work of string theory$^7$. The string scattering amplitudes
have been analysed in Ref. 8 for new symmetries that might reveal
at high energy. Furthermore the superstring Planckian scattering in
four dimensions, in the leading order eikonal, has been discussed in
Ref. 4 which has a semi-classical shock wave geometry. It is
recognised that the string theory is capable of providing answers to
some of the deep questions in quantum gravity; therefore,  it is
natural to expect that string theory will be able to predict phenomena
at the Planck scale. The results of string theory$^{4,6-8}$ and those of the
semi-classical approach$^{1-3,5}$ give similar results in the common
domains of the kinematical variables as expected.

In the recent past, Verlinde and Verlinde$^9$ have undertaken a
systematic investigation of scattering processes at Planckian energy.
Their starting point is to write the full action as the sum of
Einstein-Hilbert action for gravity and the matter part in 3+1
dimensions. It is argued that the Einstein-Hilbert (E-H)action can be
decomposed into sum of two parts, since in an appropriate
kinematic region, the small angle scattering processes are
characterized by two length scales of very different orders of
magnitude corresponding to longitudinal and transverse
directions. Therefore, it is natural to introduce two coupling
constants associated with the two parts of the gravitational
action. It was demonstrated$^9$ that one of the two actions
becomes a topological sigma model whereas the other one can be
treated as a semi-classical gravitational action. Furthermore it
was shown that the topological part of the action can be set to
zero when one considers Planckian scattering and the
semi-classical piece of the action can be expressed as a surface
term. Similarly when the masses of the in-coming particles are
negligible compared to $M_P$ the matter part of the action can
also be expressed as a surface term. Indeed, the effective
action that describes the high energy scattering is demonstrated
to have a term which is analogous to the string action and the
matter-gravity interaction part bears a strong similarity with
the vertex operators in string theory. This remarkable result
was first presented by 't Hooft in his investigations of
Planckian scattering$^5$. Verlinde and Verlinde$^9$ have derived
this result in the semi-classical approximations for quantum
gravity on a rigorous field theoretical frame-work. An
alternative geometrical description of the topological field
theory describing the Planckian scattering$^9$ has also been
discussed in Ref. 10. Recently, the semi-classical shock wave
picture of various high energy scattering processes describing
point particles, $e.g.$ point magnetic monopole and point charge
particles$^{11}$, have been discussed in literature. Furthermore a
new semi-classical approach to four dimensional Planckian energy
scattering has been proposed in Ref. 12 to calculate the
S-matrix to leading order.

In an another interesting paper$^{13}$, Verlinde and Verlinde
have studied high energy scattering processes in quantum
chromodynamics (QCD). At asymptotic energy a simple intuitive
picture emerges where the longitudinal length scales are
squeezed due to the Lorentz contraction whereas the transverse
directions remain unaffected. The final effective action
describing the high energy prcesses in QCD corresponds to a
chiral Lagrangian. Consequently, using the techniques of chiral
Lagrangian for the effective theory they were able to reproduce
several results of the high energy scattering processes in QCD.

Motivated by this semi-classical shock wave solutions of vacuum
Einstein's eq.$^{14}$, we study the high energy collision
of non-abelian gauge particles in 3+1 dimensions at Planckian
energy. However in presence of matter fields and cosmological
constant the shock wave geometry is also discussed in Ref. 15
recently. It is well known that at the Planckian energies
gravity plays a dominant role in describing interactions
among the particles. Here we present a systematic study of high
energy processes involving the gauge particles in the frame-work
of semi-classical approximation adopted in Ref. 9. Since the
coupling of the Yang-Mills to Einstein gravity is conformally
invariant at the classical level, we have to specify the gauge
fixing for the gravitational part as well as for the Yang-Mills
action. The gauge choice, ghost action and the BRST
transformations for the gravitational part have been presented
in Ref. 9.

It is shown, in the present investigation,
that the Yang-Mills ghost Lagrangian also decomposes into two parts when
one introduces two length scales in the theory. We argue that
each of the gauge fixed action and the ghost action decompose
into pieces associated with the space-time metric. Moreover the
BRST transformations reflect these effects explicitly. We
present a systematic analysis for the effective theory and write down the
vertex operators satisfying the requirements of BRST invariance
ensuring the gauge invariance of the theory.

We outline our article as follows: In section 2, we discuss the
Verlinde and Verlinde's approach to construct the topological
field theory and in section 3, we consider the Yang-Mills
coupling to the gravity in the Planckian energy limit and we show
explicitly the BRST invariance of the Yang-Mills action in
presence of gravity. The section 4, deals with the
Einstein-Hilbert action coupled to Yang-Mills in high energy
limit and we write down the canonical quantization relations.
In section 5, we discuss the semi-classical gravitational shock
wave picture for the Planckian energy scattering of non-abelian
gauge particles and we write down the S-matrix. Finally we
conclude with the discussions in section 6.

\section{Effective theory of gravity in the Planckian energy
limit}

Let us choose two light-cone coordinates $x^{\a }\equiv
(x^+, x^- )$ and two transverse coordinates $y^i \equiv (y,z)$
for $i=1,2$ with $x^{\pm} = x\pm t$ to describe the collision of
two fast moving particles with very large longitudinal momenta
in $x^{\alpha }$ plane and relatively at a large transversal
distance $(y^i)$.  The gravitational field of each of these fast
moving particles, as experienced by the other particle, takes
the form of a shock wave. Thus the interactions between the
particles will occur at the instant each particle passes through
the shock waves of the other and as a result confined to the two
dimensional transverse plane $(y^i)$.

Under a gauge choice of the metric ${\cg }_{i\a }=0$, the
metric can be written as$^9$
\be
{\cg }_{\mu\nu } = \left (\matrix {{g_{\a\b }} & {0}\cr {0} &
{h_{ij}} }\right )\nonumber \\
\ee
where $g_{\a\b }$ has the Lorentzian signature and
$h_{ij}$ is Euclidean. Notice that the Einstein-Hilbert, E-H, action
\be
S_{E-H} = -{1\over{8\pi G_N}} \int d^4x {\sqrt {-\cg }} \cal R
\ee
can be decomposed under the above gauge choice as a sum of two terms
\be
 S_{E-H} = S_L (g,h) + S_T(h,g)\nonumber\\
\ee
where
\be
 S_L (g,h) = \int d^2 x^{\a} d^2 {y^i} \sqrt g
\sqrt h ( R_h + {1\over4} h^{ij} {\partial }_i g_{\alpha\beta }
{\partial }_j g_{\gamma\delta } {\epsilon }^{\alpha\gamma}
{\epsilon }^{\beta\delta})
\ee
and
\be
S_T (h,g) = \int d^2 x^{\a} d^2 {y^i} \sqrt h
\sqrt g ( R_g + {1\over4} g^{\alpha\beta } {\partial}_{\alpha } h_{ij}
{\partial }_{\beta } h_{kl} {\epsilon }^{ik}{\epsilon }^{jl}).
\ee
If we adopt a convention, following Ref. 9, where the coordinates
donot have dimensions of length and the metric carries the dimension
of $(\rm length)^2$; we can write
\be
g_{\a\b } = {l_L}^2 {\hat g}_{\a\b },\;\;\;\; h_{ij} = {l_T}^2 {\hat
h}_{ij}
\ee
\noindent Then it is easy to see that
\bea
&&S_L ({l_L}^2 \hat g, {l_T}^2 \hat h ) = {l_L}^2 S_L(\hat g, \hat h
),\nonumber\\
&&S_T ( {l_T}^2 \hat h , {l_L}^2 \hat g) = {l_T}^2 S_T(\hat h, \hat g ).
\eea
We recall that the E-H action is always divided by the
Newton's constant $G_N={l_{p}}^2$. Now it is clear that in this
kinematical regime the E-H action splits into a strongly coupled part
$S_L$ with the coupling constant $g_L={{l_p}\over{l_L}}\approx 1$ and
a weakly coupled part $S_T$ with coupling $g_T={{l_p}\over{l_T}}\ll 1
$. It is argued$^9$ that the full E-H action can be
written as a sum of a theory to leading order in $g_T$ and a
non-perturbative theory with coupling $g_L$. Furthermore the gauge
choice ${\cal G}_{i\alpha }=0$ gives rise to ghost action which can
be written as a sum of two terms
\be
 S_{ghost} = S_L (b,c) + S_T(b,c)\nonumber\\
\ee
where
\be S_L (b,c) = \int d^2 x^{\a} d^2 {y^i} {\sqrt g}
{\sqrt h} h^{ij} b_{i\a }{\pr }_j c^{\a }
\ee
and
\be
S_T (b,c) = \int d^2 x^{\a} d^2 {y^i} {\sqrt h}
{\sqrt g} g^{\a\b } b_{i\a }{\pr }_{\b }c^i.
\ee
\noindent One can check that $S_L$ and $S_T$ scale with ${l_L}^2$ and
${l_T}^2$ respectively as in the case for $S_L(g,h)$ and $S_T(h,g)$.
We conclude that $S_L$ corresponds to a strong coupling and therefore
requires a non-perturbative treatment. On the other hand $S_T$ is in
the weak coupling regime. Consequently, the dominant configurations
contributing the partition function are  at the saddle point $S_T=0$
which corresponds to
\be
R_g =0,\;\;\;\; {\pr }_\a h_{ij} = 0.
\ee
\noindent The general solutions of eq.(11) are
\be
g_{\a\b } = {\pr }_\a X^a{\pr }_\b X^b {\n}_{ab}
\ee
\noindent where $X^a = X^a (x^\a ,y^i )$ is the diffeomorphism of
$g_{\a\b}$ to the digonal metric ${\n }_{ab}$ and the euclidean metric,
$h_{ij}$ is independent of longitudinal coordinates ($x_\a $).
Notice that Eq.(12) reminds us of the vierbein formalism and we
are tempted to make the identification $e_{\a}^a = {\pr }_{\a}
X^a $. Indeed, this identification helps us in carrying out
calculation later. Now the full action in Eq.(2) in this energy
regime can be written in terms of the boundary values ${\bar
X}^a(\tau, y^i)$ by using Stoke's theorem as
\be
S_{\pr M}=\oint dx^{\a } \int d^2y^i {\sqrt h} {\epsilon }_{ab}
(R_h {\bar X}^a {\pr}_{\a} {\bar X}^b + {\pr}_i{\bar X}^a
{\pr}_{\a}{\pr}^i {\bar X}^b ).
\ee
It has been shown in Ref. 9 that this boundary values ${\bar
X}^a(y^i)$ take arbitrary values and enable one to incorporate
matter into the theory and hence responsible for the coupling of
external matter to the gravitational field. Furthermore it has been
argued that $S_T(h,g)$ together with the ghost Lagrangian $S_T(b,c)$
describe a topological sigma model. However, the strong coupling
theory described by $S_L$, can be treated semi-classically. The
stress energy momentum tensor,
\be
T_{\a\b }= {2\over{\sqrt g}}{{\delta S_m}\over{\delta g^{\a\b }}},
\ee
\noindent is traceless as the transversal components $T_{ij}$ are
assumed to be negligible compared to the longitudianal components
$T_{\a\b}$, when one considers the Planckian energy scale.

B
\section{BRST invariance of the effective Yang-Mills action }

Consider the coupling of $SU(2)$ gauge theory, for example
W-bosons, to  gravity which is conformally invariant at the classical
level. The Yang-Mills action is
\be
S_{YM}= -{1\over{4}}\int d^4 x \sqrt {-\cg} {\cg}^{\mu\rho}
{\cg}^{\nu\l} {F^a}_{\mu\nu} {F^a}_{\rho\l}
\ee
\noindent where the field strength ${F^a}_{\mu\nu} =
\pr_{\mu}{A^a}_{\nu} - \pr_{\nu}{A^a}_{\mu} + e f^{abc}
{A^b}_{\mu}{A^c}_{\nu} $, here $A_{\mu}\equiv {A^a}_{\mu}T^a$,
being the gauge potential with $`a$' taking value in the gauge group;
$`e$' is the Yang-Mills coupling constant and $f^{abc}$ are the
structure constants evaluated when the generators, $T^a$, belong to
the adjoint representation of the group. Now with the gauge choice
(1) of the metric and the scaling arguments of $g_{\a\b}$ and
$h_{ij}$ as in Eq.(6), we write down the Eq.(15) as a sum of three terms
\bea
S_{YM}=&& -{1\over4} {{{l_T}^2}\over{{l_L}^2}}  \int d^2x^{\a}d^2y^i
\sqrt {-\hat g}\sqrt {\hat h} {\hat
g}^{\a\ga}{\hat g}^{\b\d} {F^a}_{\a\b}{F^a}_{\ga\d}\nonumber \\
&&-{1\over4} {{{l_L}^2}\over{{l_T}^2}} \int d^2x^{\a}d^2y^i
\sqrt {\hat h}\sqrt {-\hat g} {\hat
h}^{ik}{\hat h}^{jl} {F^a}_{ij}{F^a}_{kl}\nonumber \\
&&\hspace{.7in} -{1\over2} \int d^2x^{\a}d^2y^i \sqrt {-\hat g}\sqrt
{\hat h} {\hat g}^{\a\b}{\hat h}^{ij} {F^a}_{i\a}{F^a}_{j\b}.
\eea
\noindent Since the Yang-Mills coupling constant is dimensionless in
four dimensions, we show that the three terms have coefficients which
are ratios of length scales. Furthermore, the first two terms have
coefficients  (${{{l_T}^2}\over{{l_L}^2}}$) and
(${{{l_L}^2}\over{{l_T}^2}}$), where as the last term does not have a
coefficient as ratios of $l_T$ and $l_L$. The situation is analogous
to the scenario considered by Verlinde and Verlinde in the context of
QCD$^{13}$. If we identify ${{{l_L}^2}\over{{l_T}^2}}= \l $ of Ref.
13, then the Lagrangian has the exact decomposition, however we have not
incorporated the quark fields for the sake simplicity. One can argue
that the first term in (16), in this limit, is dominated by the
trivial vacuum, $i.e. {F^a}_{\a\b} =0$ so that ${A^a}_{\a}$ is a pure
gauge. In the high energy limit when
${{{l_L}^2}\over{{l_T}^2}}\rightarrow 0$, the magnetic term
(second term of Eq.(16) ) gives negligible contribution. As a result
the truncated action is represented by the last term.

So far, we have not incorporated the gauge fixing and the ghost terms
for the Yang-Mills theory. In order to present our arguments
in a transparent manner, let us consider the case of flat space. The gauge
field effective action is the sum of the Yang-Mills part, the gauge
fixed part ${\cal L}_{gf} $  and the ghost part ${\cal L}_{ghost}$,
\be
{\cal L}= {\cal L}_{YM} + {\cal L}_{gf} + {\cal L}_{ghost},
\ee
\noindent where ${\cal L}_{YM} $ is given by eq.(16) with the metric
replaced by the flat space metric. We choose Lorentz gauge to write
down the gauge fixed Lagrangian as
\be
{\cal L}_{gf} = -{1\over{2\a }}{\big ( {\pr}^{\mu }{A^a}_{\mu}\big )}^2
\ee
\noindent and the ghost part as
\be
{\cal L}_{ghost} = {\pr}^{\mu}{\n}^a {D_{\mu }}^{ab}{\omega}^b
\ee
where ${\n}^a$ and ${\omega }^b$ are the independent ghost fields,
which are in the adjoint representation of the gauge group. The
BRST transformations,
\bea
&&{\delta } {A^a}_{\mu} = -{{\e}\over{e}}D_{\mu}{\omega }^a ,\nonumber \\
&&{\delta } {\n}^a =-{{\e}\over{\a e}} {\pr }^{\mu}{A^a}_{\mu},
\nonumber \\
&&{\delta } {\omega }^a = -{{\e}\over2} f^{abc}{\omega }^b{\omega }^c.
\eea
\noindent leave the action (16) invariant, where $\e $ is
infinitesimal Grassmann parameter. Now, let us apply the
arguments of Ref. 13 for QCD to the full action (16) with a flat
metric, where the high enegy limit is implemented by the scaling
argument given by
\bea
&&x^{\a}\ra\l x^{\a}\;,\;\;\;\;\;\;\;\;\;\;\; y^i\ra y^i \; ,\nonumber \\
&&{\pr}_{\a}\ra {\l}^{-1} {\pr}_{\a}\;,\;\;\;\;\;\;\;\;\;\;\;\;
{\pr}_i\ra {\pr}_i\;,\nonumber \\
&&A_{\a}\ra {\l}^{-1}A_{\a}\;\;\;{\rm and}\;\;\;\;\; A_i\ra
A_i.\nonumber \\
\eea
Immediately, one can see that the gauge fixed and the
ghost part of the action under the above coordinate scaling can
be written as
\be
{\cal L}_{gf} = -{1\over{2\a }}{\l}^{-4}\big ({\pr}^{\a}
{A^a}_{\a}\big )^2 + \big ({\pr}^i{A^a}_i\big )^2
\ee
\noindent and the ghost part as
\be
{\cal L}_{ghost} = {\l}^{-2}{\pr}^{\a}{\n}^a {D_{\a}}^{ab} {\omega }^b
+{\pr}^i{\n }^a {D_i}^{ab} {\omega}^b.
\ee
\noindent Thus, we observe that in the high energy limit $i.e.
\l\ra 0$, the term $({F^a}_{\a\b})^2$ should be sufficiently
small and only $({F^a}_{i\a})^2$ survives. Further
${A^a}_{\a}$  being a pure gauge can be set to zero. In this
limit, the BRST transformations (20) are
\bea
&&{\delta }{A^a}_{\a} = -{{\e}\over{e}}{D_{\a}}^{ab}{\omega }^b =0,
\nonumber \\
&&{\delta }{A^a}_i =-{{\e}\over{e}} {D_i}^{ab} {\omega }^b, \nonumber \\
&&{\delta }{\n}^a =-{{\e}\over{\a e}}{\pr}^i {A^a}_i , \nonumber \\
&&{\delta }{\omega }^a = -{{\e}\over2}f^{abc}{\omega }^b{\omega }^c
\eea
\noindent and the effective Lagrangian
\be
{\cal L }_{eff}=-{1\over4} {F^a}_{i\a}{F_a}^{i\a}
\ee
\noindent is BRST invariant. It is evident that in this
approach to high energy limit the non-abelian gauge fields
behave  effectively like abelian in the flat space. Since $
{F^a}_{i\a} = {\pr}_i {A^a}_{\a} - {\pr}_{\a} {A^a}_i + e
f^{abc} {A^b}_i {A^c}_{\a}$ goes over to $ - {\pr}_{\a}
{A^a}_{i}$ , when $A_{\a}$ can be set to zero by the gauge
choice. Now we consider the case
of more general metric ${\cg }_{\mu\nu}$, in the curved spacetime. The full
action can be
written as a sum of the Yang-Mills action $S_{YM}$, gauge fixed
part $S_{gf}$ and the ghost part $S_{ghost}$.
The Yang-Mills action is given in Eq.(16), the gauge fixed part
and the ghost part can be written as
\be
S_{gf}= -{1\over{2\a }}\int d^4 x \sqrt {-\cg} {\cg}^{\mu\nu}
{\pr }_{\mu }{A^a}_{\nu }
{\cg}^{\l\rho} {\pr }_{\l }{A^a}_{\rho }
\ee
\noindent and
\be
S_{ghost}= \int d^4 x \sqrt {-\cg} {\cg}^{\mu\nu}
{\pr }_{\mu }{\eta }^a {D_{\nu }}^{ab}{\omega }^b .
\ee
\noindent Now under the scaling of the metric $g_{\a\b}$ and
$h_{ij}$ (6) the gauge fixed and the ghost part of the action can
be written as
\bea
S_{gf}=&& -{1\over{2\a }}{{{l_T}^2}\over{{l_L}^2}}
\int d^2 x^{\a} d^2 y^i \sqrt {-\hat g}\sqrt {\hat h} {\hat
g}^{\a\ga}{\hat
g}^{\b\d}{\pr}_{\a}{A^a}_{\b}{\pr}_{\ga}{A^a}_{\d}\nonumber \\
&&-{1\over{2\a }} {{{l_L}^2}\over{{l_T}^2}} \int d^2 x^{\a} d^2 y^i
\sqrt {-\hat g}\sqrt {\hat h} {\hat h}^{ik}{\hat h}^{jl}
{\pr}_i{A^a}_j{\pr}_k{A^a}_l\nonumber\\
&&-{1\over{\a }} \int d^2 x^{\a} d^2 y^i \sqrt {-\hat g}\sqrt {\hat h}
{\hat g}^{\a\b}{\hat
h}^{ij}{\pr}_{\a}{A^a}_{\b}{\pr}_i{A^a}_j
\eea
\noindent and
\be
S_{ghost}= \int d^2x^{\a}d^2{y^i} \sqrt {-\hat g}\sqrt {\hat h}
\big [ {l_T}^2 {\hat g}^{\a\b} {\pr }_{\a}{\eta }^a
{D_{\b}}^{ab}{\omega }^b + {l_L}^2 h^{ij} {\pr }_i{\eta }^a
{D_j}^{ab}{\omega }^b\big ].
\ee
\noindent In the Planckian energy limit $i.e. l_L \approx
l_p$ and $l_T \gg l_p$, since $A_{\a}$ is in a pure gauge,  as
we argued earlier ( see the discussion after Eq.(16) ). Thus
dropping the terms which have $l_{T}$ as the coefficient, the
effective action is the sum of Eqs.(16), (28) and (29)
can be written as
\be
{S^{eff}}_{YM}= -{1\over2}\int d^2x^{\pm }d^2y^i
{\sqrt {-\hat g}}\sqrt {\hat h} {\hat
g}^{\a\b}{\hat h}^{ij} {F^a}_{i\a}{F^a}_{j\b}
\ee
\noindent and the BRST transformations are
\bea
&&{\delta }{A^a}_{\a} =  -{{\e}\over{e}}{D_{\a}}^{ab} {\omega }^b =0,
\nonumber \\
&&{\delta }{A^a}_i = -{{\e}\over{e}} {D_i}^{ab} {\omega }^b ,
\nonumber \\
&&{\delta }{\n}^a = -{{\e}\over{{\a }e}}{\hat h}^{ij}
{\pr}_i{A^a}_j, \nonumber \\
&&{\delta }{\omega }^a = -{{\e }\over2}f^{abc}{\omega }^b{\omega }^c.
\eea
\noindent It is note-worthy that the Yang-Mills effective action
given in Eq.(30) respects the BRST symmetry as is evident from the
transformation (31), in the Planckian energy limit.

\section{Canonical quantization of the effective theory}

In order to have a Planckian scattering picture using the semi-classical
technique, we consider the Yang-Mills action (30) describing
non-abelian gauge particles ($e.g.$ W-bosons) of negligible mass in
comparison to $M_P$ as the external matter coupled to the E-H action (2)
\be
S_{eff} = {S^{eff}}_{E-H} + {S^{eff}}_{YM}.
\ee
\noindent These W-bosons give rise to
the energy-momentum stress tensor, which is covariantly conserved. In
the present investigation, momenta of the particles in the longitudinal
plane are of the order of Planckian energy. The transversal momenta
are negligible in comparision to $M_P$ but much larger than the
particle momenta. In this energy regime, the mass of W-bosons
are negligible and the scattering process is very well
calculable as graviton exchange between the gauge particles
dominate, with $G_{N}$ playing the role of dimensionless
coupling constant. For asymptotically large energies, all the
particle momenta are in the $x^{\a}$ direction, as a result at
the boundary $\pr M$ of the longitudinal plane M, all components
except $T_{\a\b}$ vanish. The Yang-Mills energy-momentum stress
tensor $T_{\a\b}$ can be calculated from Eq.(30) in the
Planckian energy limit ($A_{\a}$ is in pure gauge) is given by
\be
T_{\a\b} = {\hat h}^{ij}{\big
[{\pr}_{\a}A_i{\pr}_{\b}A_j - {1\over2}{\hat g}_{\a\b}
{\hat g}^{\rho\sigma}{\pr}_{\rho }A_i{\pr}_{\sigma}A_i \big ]}.
\ee
\noindent Since the in-coming and the out-going W-bosons in the
Planckian energy regime are interacting with the E-H gravity at the
boundary $\pr M$ of the longitudinal Lorentzian plane $M$, it is
convenient to represent the matter particles of a momentum flux
$P_{a\a }$, defined in terms of stress-energy tensor
$T_{\a\b}=P_{a\a}{e^a}_{\b}$, where ${e^a}_{\b}= {\pr}_{\b}X^a$
is the vierbein. Now, the momentum flux can be written as
\be
P_{a\a}= {\hat h}^{ij} {\big [{\pr}_{\a}A_i {{\pr
A_j}\over{\pr X^a}} - {1\over2} {\pr}_{\a}X^b {\n}_{ba}
{\pr}_{\rho }A_i{\pr}^{\rho }A_j \big ]}.
\ee
\noindent The variation of the Yang-Mills action (30) with respect to
$X^a $ is$^9$
\be
{\d } S_{YM} = 2{{\oint }_{\pr M}} dx^{\b }{\int d^2 y^i} {\sqrt {\hat h}}
{{\e }_{\b }}^{\a }P_{a\a }{\d }{{\bar X}^a}
\ee
\noindent where ${\bar X}^a$ take the asymptotic values of $X^a$.
Using eq.(34), we write down the effective Yang-Mills action
in terms of the gauge fields as
\bea
{{S^{eff}}_{YM}= 2{{\oint }_{\pr M}} dx^{\b } \int d^2y^i {\sqrt {\hat h}}
{{\e }_{\b }}^{\a }{\hat h}^{ij} [{\pr }_{\a }A_i {{\pr
A_j}\over{\pr {\bar X}^a}} {\bar X}^a- {1\over2} {\bar X}^a
{\pr}_{\a }{\bar X}_a  {\pr }_{\rho }A_i{\pr }^{\rho }A_j ]}
\eea

In order to determine the scattering amplitudes of W-bosons in
presence of gravity, we write down the full effective action at the
boundary $\pr M$ of the longitudinal plane by introducing
a ``time" variable $\tau $ parametrizing the coordinates
$x^{\b}(\tau )$ on the boundary. Since the boundary $\pr M$ is closed,
$\tau $ is periodic. Now from Eqs.(13) and (36) the full effective
action on the boundary $\pr M$ in (2+1) dimensions is
\bea
{S^{eff}}_{\pr M} =&& \int d\tau \int d^2y^i{\sqrt {\hat h}} {\e }_{ab}
{\pr
}_{\tau } {\bar X}^a \big ({{\bigtriangleup }_{\hat h}} - R_{\hat h}\big
){\bar X}^b \nonumber \\
&&+2\int d\tau \int d^2y^i {\sqrt{\hat h}} {{\e }_\b}^{\a}{\hat h}^{ij}
{\big [{\pr}_{\tau }A_i {{\pr A_j}\over{\pr {\bar X}^a}} {\bar X}^a-
{1\over2} {\bar X}^a {\pr}_{\tau }{\bar X}_a
{\pr}_{\rho }A_i{\pr}^{\rho }A_j \big ]}
\eea
\noindent where ${{\bigtriangleup }_{\hat h}}$ is the scalar laplacian
in the transversal $y^i$ plane. Here the action is written in terms
of the new coordinate fields ${\bar X}^a $. Note that the
time derivative of the coordinate and the gauge fields appear as
linear terms and the action is quadratic in the fields ${\bar X}^a$
as is seen in Eq.(37) . Thus the canonical conjugate momenta
corresponding to the coordinate fields ${\bar X}^a$ and the
non-abelian gauge fields $A_i$ are
\be
{\cal P}_a = {\e}_{ab} \big ({\bigtriangleup }_{\hat h} -
R_{\hat h}\big ){\bar X}^b -{\hat h}^{ij} {\bar X}_a
{\pr}_{\rho }A_i{\pr}^{\rho }A_j
\ee
\noindent and
\be
{\Pi }_j= 2{{\pr A_j}\over{\pr {\bar X}^a}} {\bar X}^a
\ee
\noindent by considering the parameter $\tau $ as the quantization
time variable. The canonical conjugate momenta $P_a$ thus derived is
invariant under the BRST transformations given by eq.(31), since
$T_{\a\b }$ is also BRST invariant. In the absence of gauge fields $(
A_i = 0 )$ the canonical quantization of ${\bar X}^a$ in the 2+1 is
discussed in Ref. 9 and the equal time commutation relations become
\be
\big [{\bar X}^b(y_1) , {\cal P}_b(y_2)  \big ] = i {\delta }^{(2)}
(y_1, y_2 ).
\ee
\noindent However in the presence of non-abelian gauge fields the
equal time commutator for the coordinate fields can be written as
\be
\big [{\bar X}^a(y_1) , {\bar X}^b(y_2)\big ] = i {\e }^{ab} {\bar
f}{(y_1, y_2 ) }\; ,
\ee
\noindent where ${\bar f}(\big(y_1, y_2\big )$ is the Green function
for the interacting W-bosons in presence of gravity. Now we write
down classical equations of motion from Eq.(37) by taking the variation
with respect to the coordinate field ${\bar X}^a$ as
\be
{\pr }_{\tau } {\bar X}^a \big ({\bigtriangleup }_{\hat h} - R_{\hat
h}\big ) = {\e }^{ab}P_{a,\tau }
\ee
where
\be
P_{a,\tau } =-2{{\e}_\b}^{\a}{\hat h}^{ij}
\big [{\pr}_{\tau }A_i {{\pr A_j}\over{\pr {\bar X}^a}}
-{1\over2}{\pr}_{\tau }{\bar X}_a {\pr}_{\rho }A_i{\pr}^{\rho }A_j \big ].
\ee
\noindent This expression (42) represents the Einstein equation
in presence of Yang-Mills source at the Planckian energy and relates
coordinate fields to their canonical conjugate momenta.

\section{ Semi-classical picture of scattering}

Now in order to have an intuitive semi-classical
picture we consider two W-bosons scattering process.
The momentum flux $P_{a,\tau }$ for the W-bosons is the quantum
mechanical operator in the Hilbert space of the gauge theory and
is divided into left and right movers with momenta ${p_-}^i$ and
${p_+}^j$ respectively as W-boson masses are assumed to be
negligible compared to the $M_P$. As a result the boundary $\pr
M$ is divided into four different asymptotic regions of
B
space-time (${\cal I}^+ $ and ${\cal I}^-$).
Considering the in-coming momentum flux at asymptotic
past ${\cal I}^- $ in the following form
\bea
&&{P_-}^{in}(x,y) = {p^-}_1- \delta (x_+ - x_{1+}) \delta
(y-y_1),\nonumber \\
&&{P_+}^{in}(x,y) = {p^+}_2 + \delta (x_- - x_{2-}) {\delta
(y-y_2)}
\eea
where ${p^-}_1$ and ${p^+}_2$ are the longitudinal momentum of the
B
in-coming left  and right moving W-bosons in ${x^-}_1$ and ${x^+}_2$
directions respectively. The dynamics is assured by the $\delta
$-functions in the Eq.(44). Now the classical solutions of the
reduced Einstein Eq.(42) are given by
\bea
&&{X_-}^{in} = x^- - {p_1}^- {\theta }(x_+ - x_{1+}) {\bar f}(y_1,
y_2),\nonumber \\
&&{X_+}^{in}= x^+ -{p_2}^+ {\theta }(x_- - x_{2-}) {\bar f}{(y_1,
y_2)}.
\eea
\noindent The classical field configurations (45) exhibit a
discontinuity known as gravitational shock wave at the $x^{\a}$
trajectory of the particles are incompartible according to the commutation
relations (41) and may give rise to the quantum mechanical effect of the
gravitational shock waves. Thus in the quantum mechanical description
it might be possible to consider two colliding shock waves. However in this
frame-work, two W-bosons scattering by
the exchange of gravitons has a semi-classical picture of
gravitational shock wave colliding with a slow moving W-boson. This
two particle scattering amplitude has been derived in an elegant way
way by 't Hooft$^1$, considering the effect of shock wave of one of
the two colliding particles on the wave packet of the other.

In order to determine S-matrix for the W-bosons scattering due to
the exchange of gravitons, we write down the effective Yang-Mills
coupling to Einstein gravity at Planckian energy, using Eqs.(34) and
(37) in terms of ``vertex operators" defined$^9$ as
\bea
V(P) = P exp{\big [2i \int d\tau \int d^2 y^i {\sqrt {\hat h}} P_{a,\tau }
{(\tau ,y) {\bar X}^a (y)}\big ]}.
\eea
Here `P' represents the path order product, as the ${{\bar X}^a(y)}$
do not mutually commute. It is note-worthy that the vertex operators
(46) do satisfy the BRST invariance (31). Furthermore the presence of
four asymptotic regions of space-time correspond to four vertex
operators describing in-coming and out-going momentum flux at
${\cal I}^-$ and ${\cal I}^+$. A brief study of the scattering
amplitude for scalar particles has been discussed in Ref.[9]. Now, we
write down the four point vertex function representing the
scattering amplitude for the W-bosons  which may be computed by
calculating the expectation value of the vertex operators
\bea
&& <V({P_+}^{in})V({P_-}^{in})V({P_+}^{out})V({P_-}^{out})>\nonumber
\\
&& =exp\big [i\int d^2y_1 {\sqrt {\hat h}}\int d^2y_2 {\sqrt {\hat h}}
{{P_+}^{in} (y_1) {\bar f}(y_1, y_2)} {{P_-}^{in}(y_2)} \big ].
\eea
\noindent The mutually commuting operators ${{P_{\pm}}^{in/out}(y)}$
can be used to characterize the in-coming and out-going states of
W-bosons. Furthermore by momentum conservation of
out-flux and in-flux the scattering amplitude can be calculated using
Eq.(44). In this approach$^9$ the general covariance is broken in the
transverse $y^i$ direction. However in some physical situations the
general covariance remains unbroken, $e.g.$ the flat transversal
metric ($h_{ij}={\delta }_{ij}$) and analogy with string amplitude is
very close. In this special case of flat transversal metric, the
string amplitude calculation may be carried out giving rise to the
scattering amplitudes. It is shown in Ref. 5 that the S-matrix
element can also be written in a form similar to the string
amplitude. This can be argued as, at distances much smaller than
Planck length, one requires an explicit cut off because of the
difficulty of renormalizability of quantum theory of Einstein
gravity as a result it is believed that in this regime the
conventional field theory should be replaced by string theory.
Further 't Hooft's proposal$^5$ for black hole S-matrix is
analogous to the expression (47), apart from the fact that the
transversal coordinates are replaced by angular coordinates.
However in this semi-classical approach, it is not possible to
consider angular coordinates in the transversal palne, as in the
limit radial coordinate $r\ra 0$, the momentum transfer in the
transverse plane is also large.

\section{Discussions}

To conclude, in this article we have studied small angle scattering
of non-abelian gauge particles in the presence of nontrivial gravity.
We have shown that in the Planckian energy limit, the gauge particles
described by the Yang-Mills action is gauge invariant even in the absence
of ghost and gauge fixed terms. The relevent gravitational modes
mediating between the gauge particles at Planck energy are described
by means of a topological field theory and the dynamics is described
by the boundary values of $X^a$. The scattering amplitude can be
calculated explicitly for the non-abelian gauge particles which is
valid only to leading order in transverse coupling constant $g_T$ and
is non-perturbative in the longitudinal coupling constant $g_L$.
It would be interesting to consider the high energy scattering
of the particles that appear as massless excitations of string
theory and which might acquire masses due to various mechanisms
in this frame-work$^9$. Since the string effective action is
endowed with a rich symmetry structure, we expect that the high
energy scattering amplitudes derived from such an action will
have several novel features. We hope to address these issues in
future.

\bigskip
\bigskip

\noindent {\Large {\bf Acknowledgements}}

\bigskip

Authors, (SK) would like to thank T. Jayaraman and T. Sarkar for
stimulating discussions and (JM) would like to thank the Newton
Institute for Mathematical Sciences for warm hospitality where
this work was completed.

\bigskip

\def\np{{\it Nucl. Phys.}\ {\bf B}}
\def\pl{{\it Phys. Lett.}\ {\bf B}}
\def\prd{{\it Phy. Rev.}\ {\bf D}}
\def\prl{{\it Phys. Rev. Lett.}}
\def\ijmp{{\it Int. J. Mod. Phys.}\ {\bf A}}
\def\ml{{\it Mod. Phys. Lett.}\ {\bf A}}

\vfil
\eject
\end{document}